\documentclass[sigconf]{acmart}
\usepackage{wrapfig}
\AtBeginDocument{%
  \providecommand\BibTeX{{%
    \normalfont B\kern-0.5em{\scshape i\kern-0.25em b}\kern-0.8em\TeX}}}

\setcopyright{acmcopyright}
\begin{document}

\title{\emph{PRI-Attack:} Person Re-identification Attack on Wearable Sensing}

\author{Mohammad Arif Ul Alam}
\email{mohammadariful\_alam@uml.edu}
\affiliation{%
  \institution{University of Massachusetts Lowell}
  \state{MA}
  \country{USA}
  \postcode{01201}
}


\begin{abstract}
Person re-identification is a critical privacy attack in publicly shared healthcare data as per Health Insurance Portability and Accountability Act (HIPAA) privacy rule. In this paper, we investigate the possibility of a new type of privacy attack, Person Re-identification Attack (\emph{PRI-attack}) on publicly shared privacy insensitive wearable data. We investigate user's specific biometric signature in terms of two contextual biometric traits, physiological (photoplethysmography and electrodermal activity) and physical (accelerometer) contexts. In this regard, we develop a Multi-Modal Siamese Convolutional Neural Network (\emph{mmSNN}) model. The framework learns the spatial and temporal information individually and combines them together in a modified weighted cost with an objective of predicting a person's identity. We evaluated our proposed model using real-time collected data from 3 collected datasets and one publicly available dataset. Our proposed framework shows that PPG-based breathing rate and heart rate in conjunction with hand gesture contexts can be utilized by attackers to re-identify user's identity (max. 71$\pm$3) from HIPAA compliant wearable data. Given publicly placed camera can estimate heart rate and breathing rate along with hand gestures remotely, person re-identification using them imposes a significant threat to future HIPAA compliant server which requires a better encryption method to store wearable healthcare data.
\end{abstract}




\maketitle

\section{Introduction}


\begin{figure}
  \centering
  \includegraphics[width=0.6\linewidth]{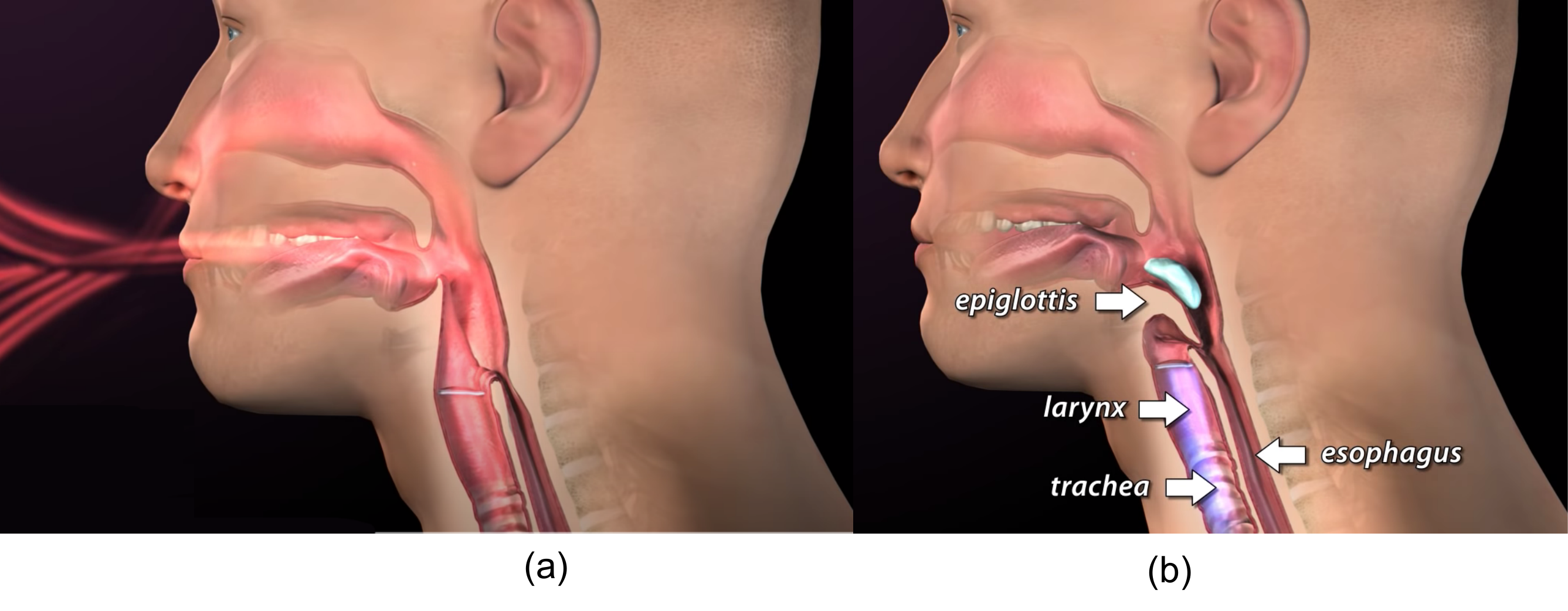}
  \vspace{-.2in}
  \caption{(a) Normal Breathing involve mouth and nose based inhalation, (b) During drinking activity, inhalation follows a different pattern due the pressure created by larynx which causes epiglottis to block the air pass through trachea.}
  \label{fig:breathing_eating}
  \vspace{-.4in}
\end{figure}

Due to the immense necessity of monitoring breathing rate and heart rate continuously, researchers developed efficient privacy insensitive commodity sensor technologies and sophisticated machine learning techniques for monitoring photoplethysmography (PPG) and electrodermal activity (EDA) using wearables \cite{empatica}. To progress research productivity, researchers often share raw sensor data and Protected Health Information (PHI) in public server with appropriate Covered Entity (CE), encoding and de-identification according to HIPAA Privacy Rule \cite{hipaa}. As per CE of HIPAA Privacy Rule \cite{hipaa}, raw PPG, EDA, Skin Temperature and accelerometer (ACC) data are not privacy sensitive and do not require any encoding while sharing publicly. On the other hand, recent evolution of face and remote PPG detection techniques using public camera recorded video data \cite{rppg1,rppg2,rppg3} raise important privacy concern: {\it Does physiological sensing (PPG, EDA, Temp) carry any user specific biometric signature?}

Many researchers showed PPG can be utilized as an additional biometric signature along with user's personal identifiers for personal device authentication purpose \cite{auth_ppg1, auth_ppg2, auth_ppg3}. Though, user's personal identifier has never been shared along with PHI wearable data, the user authentication method can not be used as an attack over user's privacy in publicly available HIPAA compliant wearable data. However, recent days, ubiquitous computing researchers provide efficient quantification of physiological sensing (say heart rate) and contextual information (physical activities) to provide healthcare services including stroke prevention, epileptic seizure detection etc \cite{acc_ppg1,acc_ppg2,acc_ppg3, acc_ppg4, acc_ppg5}. This additional contextual information has been widely shared publicly as HIPAA privacy rule does not forbid in sharing such data in public. In this paper, we argue that, the quantification of simultaneously collected multi-modal sensor aided physical and physiological contexts carry unique biometric signature of individuals that may involve a new type of potential privacy attack, Person Re-identification Attack (\emph{PRI-attack}), which can be a serious privacy threat to billions of publicly shared healthcare data.

In this paper, we investigate the potential privacy threat occurred on privacy preserving activity and physiological (PPG, EDA, Temp) sensing data. The central {\bf hypotheses} of this paper is that, {\it each microscoping human activity has person specific unique physiological response which can reveal user's identity via a multi-modal deep Siamese Neural Network} \cite{mSNN}. As an example, Fig \ref{fig:breathing_eating} shows the normal and drinking water activity behaviors in terms of breathing. It has been clearly depicted that drinking water activity responses on breathing pattern has dissimilarities to the normal breathing pattern. We argue that, inhalation and exhalation pattern while performing individual activity are unique biometric signatures that can be learned and used to extract user's identity. It should be depicted that drinking water activity can be detected accurately using both surveillance camera as well as wrist worn ACC accurately.

In this regard, we develop a multi-modal Siamese Neural Network (\emph{mmSNN}), that provides the following {\bf key contributions}

\begin{figure*}[!htb]
\begin{minipage}{0.3\textwidth}
\begin{center}
  \includegraphics[width=0.8\linewidth]{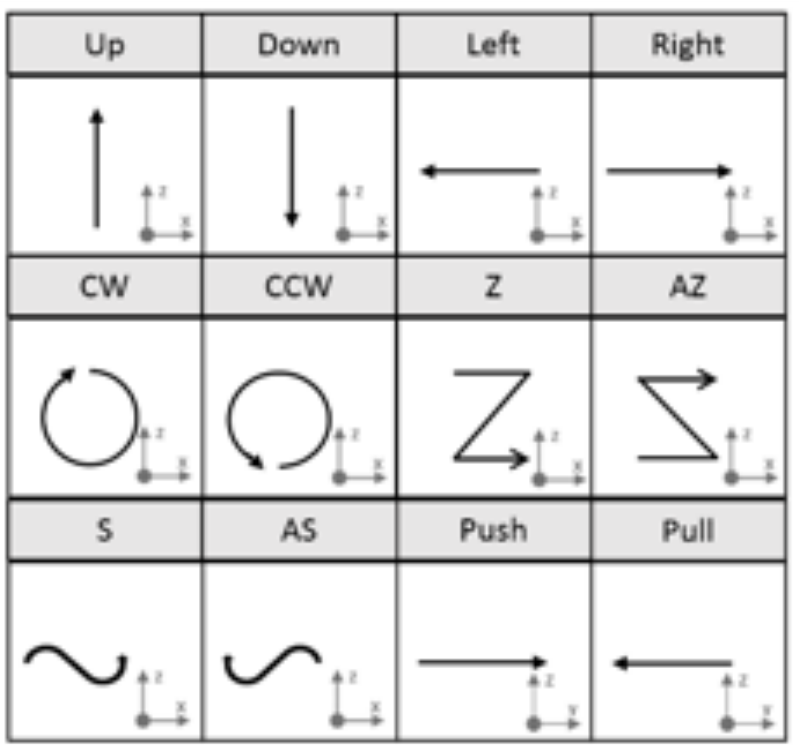}
  \caption{Hand gesture dictionary considering the wearable watch has been worn in the wrist of user's dominant hand.}
  \label{fig:gesture_dictionary}
\end{center}
\begin{center}
  \includegraphics[width=\linewidth]{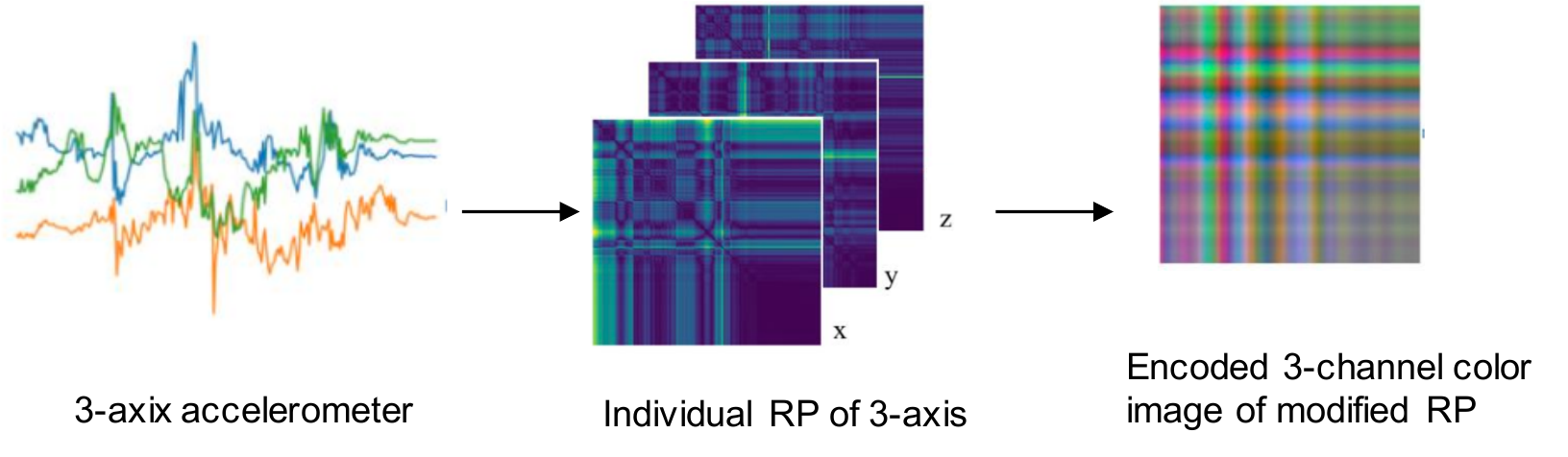}
  \caption{Modified Recurrent Plot (RP) encoding of 3-axis ACC signal into single 3-channel color image representation}
  \label{fig:acc_encode}
\end{center}
\end{minipage}
\begin{minipage}{0.2\textwidth}
\end{minipage}
\begin{minipage}{0.67\textwidth}
\begin{center}
  \centering
  \includegraphics[width=\linewidth]{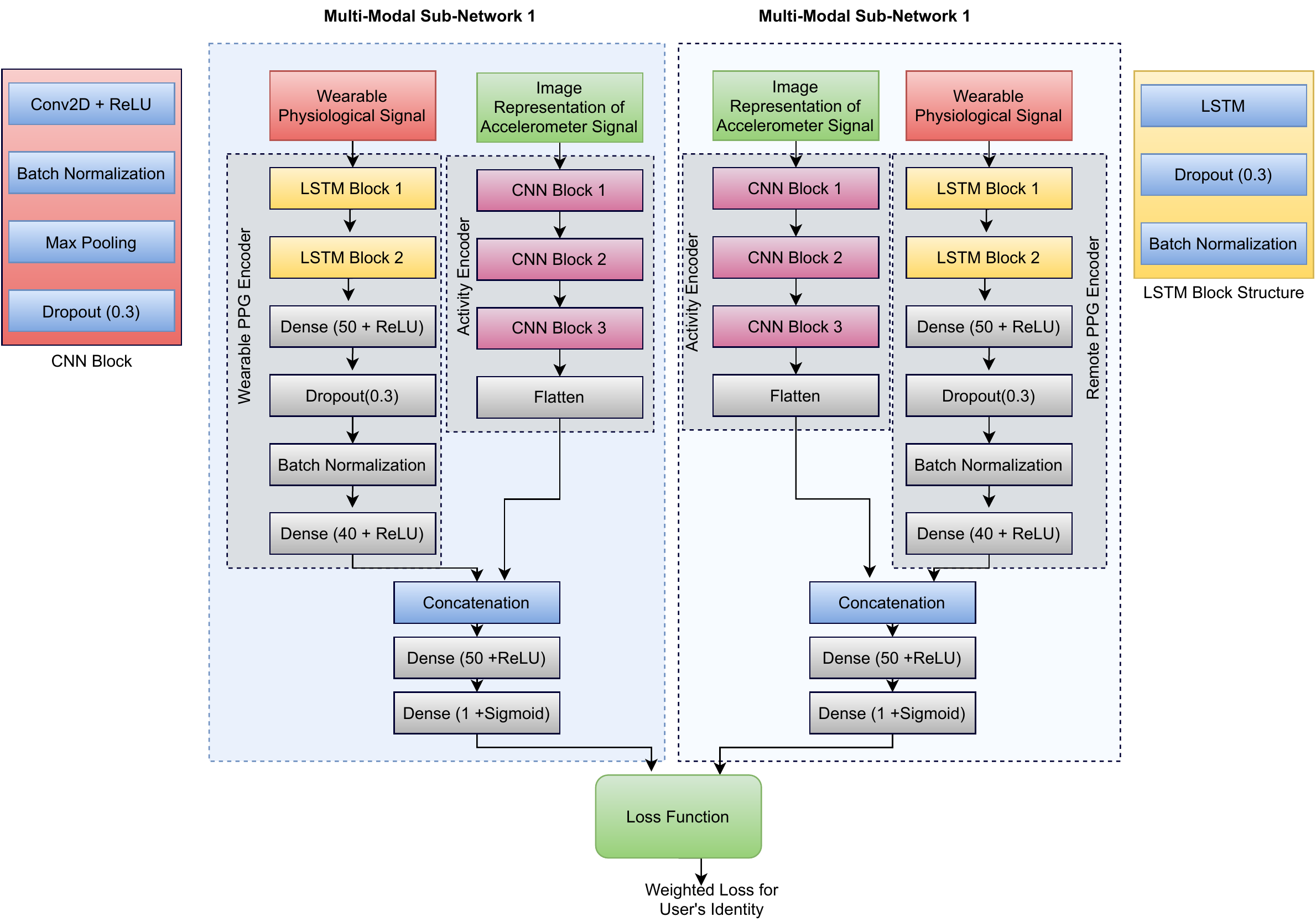}
  \vspace{-.2in}
  \caption{Schematic diagram of our proposed Multi-Modal Siamese Neural Network \emph{mmSNN} where the goal of learning the feature representation is to minimize the contrastive loss in case for corresponding data and maximize the contrastive loss otherwise.}
  \label{fig:mmSNN}
\end{center}
\end{minipage}

\end{figure*}

\begin{itemize}
\item We develop a hand gesture recognition and localization framework from wearable ACC data and develop an image representation of ACC data within detected gesture activity window.

\item We develop a Multi-Modal Siamese Neural Network (\emph{mmSNN}) that consists of two sub-networks. \emph{mmSNN} leverages a spatio-temporal architecture to generate multi-modal encodings of time series physiological sensing and image representation of ACC data within the gesture activity window.

\item We develop an efficient weighted cost function on two sub-networks together as a decision function to find the matching of user's identity.

\item Finally, we evaluate our proposed \emph{PRI-attack} framework on three in-house collected data and one publicly available data.
\end{itemize}

\section{Hand Gesture Feature Learning }

\subsection{ Gesture Recognition and Localization}
We use GestureKeeper (Implementation Code: \cite{gesture_keeper_code}) technique which is a robust hand-gesture recognition system based on a wearable 3-axis ACC sensor signals \cite{gesture_keeper}. In this regard, we first define 12 different hand gesture movements as shown in Fig~\ref{fig:gesture_dictionary}. Here, the hand gestures are as follows: {\bf Up}: Moving hand straight bottom to up, {\bf Down}: Moving hand straight up to down, {\bf Left}: Moving hand horizontally straight from right to left, {\bf Right}: Moving hand horizontally straight from left to right, {\bf CW}: Rotating hand clock-wise, {\bf CCW}: Rotating hand anti-clock-wise, {\bf Z}: Z trajectory starting from above, {\bf AZ}: Mirror Z trajectory starting from below, {\bf S}: Waving to right, {\bf AS}: Waving to left, {\bf Push}: Moving away horizontally from body, {\bf Pull}: Moving towards body horizontally. We first detect the start of a gesture recorded in ACC data stream using a recurrence quantification analysis (RQA) technique. RQA enables the detection of transitions in the dynamical behavior (e.g. deterministic, chaotic, etc.) of the observed system. We select a window size of 80 samples with 80\% overlap between consecutive windows, which represents 2.5 seconds of ACC data (32 Hz frequency), sufficient for capturing even the longest of gestures we selected. After detecting the starting point of hand gesture, we detect the gesture which is a 12-class problem. We use support vector machine (SVM) classifier using the radial kernel. As input features of the SVM, we use both statistical features and samples of the acceleration signal features (mean, root mean square (RMS), median, variance, standard deviation, skewness and kurtosis, angular velocity, \# of the 3D acceleration, and magnetism time series). We re-sample the features using x, y and z-axis acceleration of raw ACC sensor, thus the final set of features consists of 12 statistical features and number of samples of the re-sampled acceleration signal over each x-, y-, and z-axis time series of ACC. Finally, we train SVM algorithm using the above features for identifying gestures as well as localizing (start and end point of each detected gesture) them in the time window.


\subsection{Image Representation of ACC Data Within Gesture Window}
After localizing start and end point of a recognized gesture from hand worn ACC, we encode entire gesture window related ACC signal as images \cite{encode_acc} (Implementation Codes: \cite{encode_acc_code}). In this regard, as shown in Fig \ref{fig:acc_encode}, at first, we consider Recurrent Plot (RP) of each ACC axis (3-axis) data. RP is a visualization method to analyze nonlinear data points on phase space trajectories of a dynamic time-series system, whose states are typically in a rather intricate manner \cite{rp1, rp2}. We utilize RP to encode 3-axis signals as RGB channels of color images so that their correlation information can be encoded. In this regard, we use Lu et. al. proposed modified RP encoding method to generate 3-channel color image from 3-axis ACC signal using a modified RP matrix construction method \cite{encode_acc} (Fig \ref{fig:acc_encode}).

\section{Multi-Modal Siamese Neural Network (\emph{mmSNN})}
Siamese Neural Network (SNN) consists of two similar sub-neural networks with identical parameters and shared weights \cite{siamese}. These two sub-neural networks are connected by a contrastive loss function, which calculates a similarity score between two input samples based on the feature vector representation of the two sub-neural networks. \emph{PRI-attack} extends SNN to Multi-Modal sub-neural networks where each sub-network consists of a CNN-based ACC image encoder and a LSTM-based physiological sensor (PPG/EDA/Temp) encoder. Both of the encoders (CNN and LSTM) are concatenated together to generate a multi-modal fused feature representation (Fig~\ref{fig:mmSNN}). Finally, two identical sub-networks are fused together using a weighted cost function.
\subsection{Multi-modal Feature Representation}
\emph{mmSNN} takes a pair of multi-modal inputs within a detected gesture window: (i) a ACC image and (ii) a time-series of physiological signal. To achieve maximized similarities between two embeddings, a Siamese Network was designed that (1) learns the joint embeddings between the physiological signal and ACC images using deep encoders and (2) maximizes the similarity measure between the two modalities. Let a multimodal dataset $X=\{ p_g,a_g \}$ where $p$ and $a$ represent $g^{th}$-gesture related physiological and ACC image. The objective is to build two encoders that convert the physiological signal and ACC image into a common space, $\eta$. The physiological data encoder ($\phi$) that maps the physiological signal into the common space was built using LSTM modules of different sizes to facilitate learning of temporal features from the physiological data. A linear layer was then applied to project the learned features into a one-dimensional feature vector in the common space, $\eta$. Similarly, the ACC image encoder, $\alpha$, mapped the input image into the common space ($\eta$) using Convolutional Neural Networks (CNN) with different sizes of convolutional layers. Again, a linear layer was applied to produce a one-dimensional feature vector in the common space ($\eta$). The output of two encoders ($\phi$ and $\alpha$) was then concatenated using a compatibility function ($ F(p,a)=[ \phi (p), \alpha (a) ] $). $F$ is used to measure the similarity between two sample pair embeddings in the common space ($\eta$). Finally, the output of $F$ from two subnetworks is passed to a weighted loss function for joint identification and verification. We omit physiological and ACC signal encoders (shown in Fig~\ref{fig:mmSNN}) description due to space constraints.

\subsection{Weighted Jointed Identification}
During the training process, we build a joint identification and verification approach \cite{41} to define our training objective where softmax loss function has been used to compute identification cost which has been integrated into our final training objective function defined as follows
\begin{equation}
V(\eta) = P(q=c|\eta) = \frac{exp(W_c\eta)}{\sum_k{exp(W_k\eta)}}
\end{equation}
$\eta$ is the common feature vector output from multi-modal feature learning step and $q$ is the identity. $W_c$ and $W_k$ indicate the $c^{th}$ and the $k^{th}$ column of the softmax matrix $W$, respectively. $W$ is the feature weight matrix of common feature vector $\eta$. The weighted joint identification objective function, which is used for training, incorporates the ability to predict a persons identity.

\begin{table*}[!h]
  \begin{center}
\begin{small}
  \caption{Comparison of \emph{PRI-attack} performance on each dataset (D1, D2, D3, D4) of re-identifying user's identity using wearable based hand-gesture aware physiological sensor signals: Photoplethysmograph (PPG), Electrodermal Activity (EDA), Skin Temperature (Temp), EDA, Heart Rate (HR) derived from PPG signal \cite{hr_ppg}), Breathing Rate (BR) derived from PPG \cite{br_ppg}), BVP (Blood Volume Pulse derived from PPG), IBI (Inter Beat Interval derived from PPG), Tonic Component (TC) of EDA signal \cite{eda_feat}) and Phasic Component (PC) of EDA data \cite{eda_feat}). D4 has no EDA data}
  \label{tab:result-table}

  \begin{tabular}{|p{0.9cm}|p{1.45cm}|p{1.45cm}|p{1.45cm}|p{1.45cm}|p{1.45cm}|p{1.45cm}|p{1.45cm}|p{1.45cm}|p{1.45cm}|p{1.45cm}|}
      \hline
 {\bf Data \#Users}     & {\bf PPG}  & {\bf HR} & {\bf BR} & {\bf BVP}& {\bf IBI}  & {\bf EDA } & {\bf TC } & {\bf PC } & {\bf Temp}  
    \\ \hline
    D1 (5) & $69.65\pm5.4$  & $70.45\pm2.4$ & {\bf 71.58$\pm$3.1} & $66.65\pm4.6$  & $67.38\pm3.8$ & $52.43\pm9.4$ & $53.42\pm8.9$  & $54.97\pm7.6$ & $49.23\pm8.2$    \\ \hline
    D2 (8) & $67.12\pm5.1$  & $69.29\pm3.5$ & {\bf 70.54$\pm$2.9} & $65.86\pm3.9$  & $65.43\pm3.1$ & $50.87\pm9.1$ & $50.38\pm9.1$  & $51.64\pm8.1$ & $48.43\pm7.2$    \\ \hline
    D3 (22) & $66.44\pm6.6$  & {\bf 67.38$\pm$4.7} & $67.16\pm5.1$ & $64.95\pm4.5$  & $64.88\pm4.9$ & $50.63\pm9.6$ & $51.75\pm8.4$  & $50.76\pm7.5$ & $45.5\pm11.4$    \\ \hline
    D4 (28) & $65.65\pm7.1$  & {\bf 66.95$\pm$4.3} & $66.85\pm4.1$ & $65.82\pm6.1$  & $66.75\pm5.3$ & &   &  & $34.8\pm12.5$    \\ \hline

  \end{tabular}
\end{small}
  \end{center}
\end{table*}
\section{Experimental Evaluation}
\subsection{Datasets}
We use two datasets to evaluate \emph{PRI-attack} model performance which are described as follows:

{\bf D1: Gamer's Fatigue Dataset}: We collected data on 5 recruited student video games players (age ranges from 19-25) for 7 days while wearing Empatica E4 watch \cite{empatica} through the appropriate institutional IRB approval \footnote{Previously published work references with IRB number will be provided upon acceptance}. Empatica E4 watch consists of ACC (ACC), electrodermal context (EDA), photoplethysmography (PPG) and skin temperature (TEMP) sensors.

 {\bf D2: Restaurant Data}: We recruited 8 student volunteers to perform a preparing and eating sandwich activity in restaurant environment for 20 minutes each while wearing Empatica E4 watch with IRB approval \cite{empatica}${}^1$ . Empatica E4 watch consists of ACC (ACC), electrodermal context (EDA), photoplethysmography (PPG) and skin temperature (TEMP) sensors.

{\bf D3: Older Adults Data}: We collected data on 22 recruited older adults (75-95 years of old) to perform 13 scripted activities while wearing Empatica E4 watch with appropriate IRB approval \cite{empatica}${}^1$.

{\bf D4: Healthy Adults Fatigue Dataset}: We have used publicly available health adults fatigue dataset \cite{luca20,luo20} collected from 28 healthy individuals (26–55 years of age, average age 42 years, 41/51\% female/male) from 1 to 219 consecutive days (mean 35, median 9, total 973 days) while wearing multisensor wearable device, Everion (Biovotion AG, Switzerland \cite{biovotion}) over a 1-week period. The device combines a 3-axis ACC, barometer, galvanic skin response electrode, and temperature and photo sensors (see \cite{luca20,luo20} for more details). 
\begin{wrapfigure}{R}{5cm}
\centering
  \includegraphics[width=\linewidth]{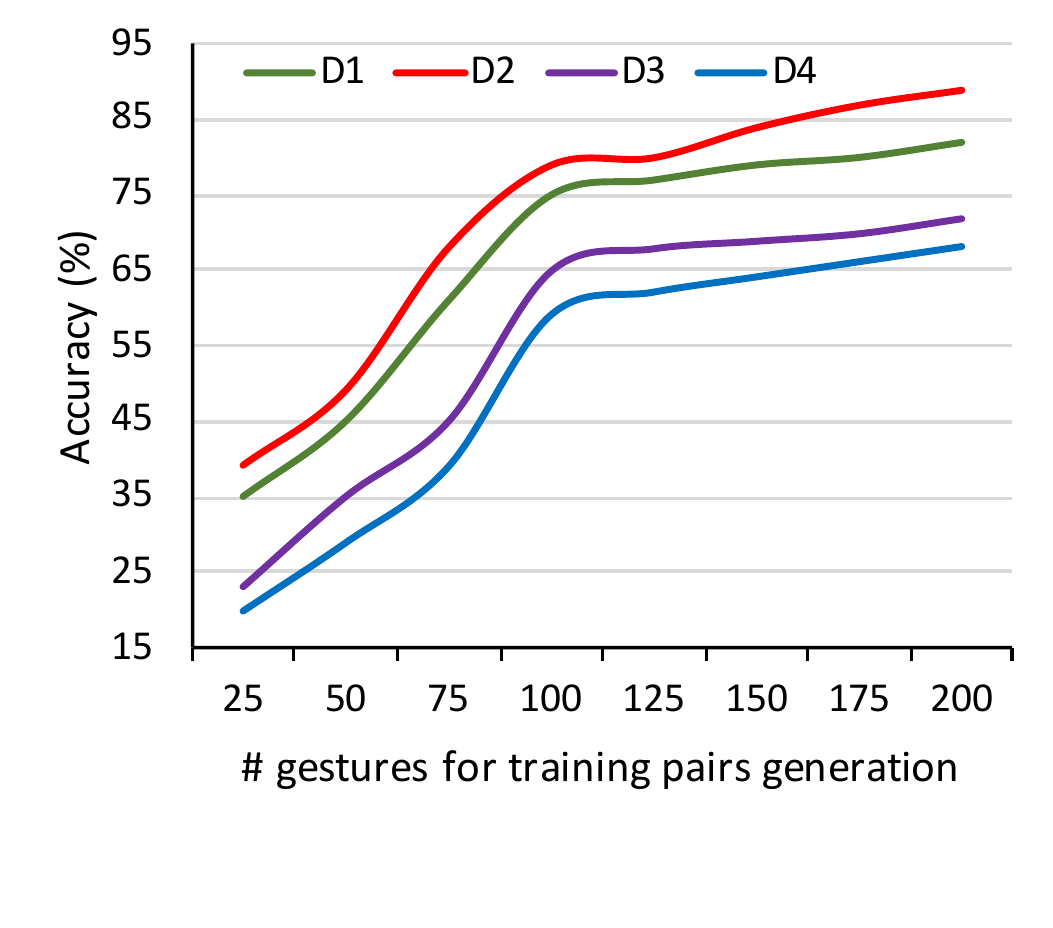}
  \caption{User's re-identification accuracy changes over \# of gesture windows in the training sample for PPG signal}
  \label{fig:results}
\end{wrapfigure}
\subsection{Data Preparation}
To generate appropriate input to \emph{mmSNN}, we need pairs of ACC and physiological signal windows that are similar, and, pairs of ACC and physiological signal windows that are dissimilar. The objective function $F$ will minimize the loss while similar pairs have been found and maximize loss otherwise. The similar pairs will be associated with the user's identity in the training set. As stated in our {\bf hypotheses}, we consider our feature space vector $\eta_i$ will be similar to another $\eta_j$ if both belongs to same individuals given the same gesture, i.e., the gesture activity responses on physiological-gesture combined feature representation is a unique biometric for each individual. Considering the above hypothesis, first, we use pre-trained GestureKeeper \cite{gesture_keeper_code,gesture_keeper} model and used it to detect and localize gesture window. During the training of \emph{mmSNN}, we consider user's identity of the training samples are known. We converted ACC signal to image and took physiological signal (sequential data) within the gesture window and train \emph{mmSNN} model along with the identity function. ACC image will ba passing through the ACC encoder and physiological signal will be passing through the physiological encoder while training \emph{mmSNN} model for similarity/dissimilarity learning based user's identity classifier. We consider $accuracy = \frac{TP+TN}{TP+TN+FP+FN}$ and Standard Deviation $\pm\%$ as evaluation metrics for re-identifying users for all gesture window in the testing data. Since, our framework is the first of its kind, we could not compare with any other frameworks with our results.

\subsection{Results Analysis}
We investigate different physiological signals i.e., Photoplethysmograph (PPG), Electrodermal Activity (EDA), Skin Temperature (Temp), EDA, Heart Rate (HR) derived from PPG signal \cite{hr_ppg}), Breathing Rate (BR) derived from PPG \cite{br_ppg}), BVP (Blood Volume Pulse derived from PPG), IBI (Inter Beat Interval derived from PPG), Tonic Component (TC) of EDA signal \cite{eda_feat}) and Phasic Component (PC) of EDA data \cite{eda_feat}) signals' privacy vulnerabilities using our proposed method. Table \ref{tab:result-table} presents the details results of our investigations on different databases considering randomly selected 100 episodes of detected gestures as our training candidates which have been used to generate similar and dissimilar data pairs to train our core \emph{mmSNN} model. The results show that PPG signals based \emph{mmSNN} algorithms provide high accuracy (65\%+) for all datasets which confirms that user's gesture specific PPG signal provides unique biometric signature that can be used against identity theft. On the other hand, skin temperature (Temp) and EDA signals derived physiological features are not that much significant (<55\%) in detecting user's identity. While analyzing more details, we found that HR and BR provides higher user specific biometric signature (>66\%+) which can be derived from PPG signal. Also, we can see that, as the number of users increases (\# users are 5, 8, 22 and 28 in datasets D1, D2, D3 and D4 respectively), the user's identity revealing capabilities are decreasing. Fig~\ref{fig:results} shows accuracy comparisons on different datasets in terms of the number of gesture windows in the training sample which shows that as the training samples are getting increased, the capabilities of re-identifying users are also increased for PPG signal on each dataset.

\section{Conclusion}
The evolution of ubiquitous sensing technologies has enabled modern intelligent systems to monitor physical (such as cooking, walking, gestures) and physiological (such as heart rate, breathing rate) contexts more accurately than ever in privacy preserving way. While the wearable and ambient sensing technologies in conjunction with complex deep learning models are getting more powerful in detecting microscoping human contexts, the data privacy has become a serious concern given the easy availability of multiple technologies. The Health Insurance Portability and Accountability Act (HIPAA) \cite{hipaa} states no clear indication of storing de-identified multi-source wearable sensor-based health contextual data in a single location (HIPAA compliant server). In this paper, we argue that, contextual responses of mult-source healthcare data can easily be utilized to perform a person re-identification attack if the attacker can learn few gesture related physiological features of the individual. While modern computer vision technology can be easily utilized to learn hand gestures and corresponding physiological signal (heart rate, breathing rate) from public surveillance camera, these huge amount of recorded videos can be easily utilized by the attackers to learn user specific biometrics to reveal identity from HIPPA compliant serve stored wearable sensing data. 
\section{Acknowledgment}
This work is supported by an Internal Seed grant of the authors affiliated university


\begin{thebibliography}{00}
\bibitem{hipaa}Atchinson, Brian K.; Fox, Daniel M. (May–June 1997). "The Politics Of The Health Insurance Portability And Accountability Act" . Health Affairs. 16 (3): 146–150. doi:10.1377/hlthaff.16.3.146
\bibitem{who19} Burn-out an "occupational phenomenon": International Classification of Diseases, 28 May 2019 Departmental news, https://www.who.int/news/item/28-05-2019-burn-out-an-occupational-phenomenon-international-classification-of-diseases 
\bibitem{walker20}Walker, W.H., Walton, J.C., DeVries, A.C. et al. Circadian rhythm disruption and mental health. Transl Psychiatry 10, 28 (2020). https://doi.org/10.1038/s41398-020-0694-0
\bibitem{bai20}Yang Bai, Yu Guan, Wan-Fai Ng: Fatigue assessment using ECG and actigraphy sensors. ISWC 2020
\bibitem{empatica}Empatica E4 Watch: https://www.empatica.com/research/e4/ 
\bibitem{sss}Hoddes E. (1972). "The development and use of the stanford sleepiness scale (SSS)". Psychophysiology. 9 (150).
\bibitem{sleep2peak}Brunet JF, Dagenais D, Therrien M, Gartenberg D, Forest G. Validation of sleep-2-Peak: A smartphone application that can detect fatigue-related changes in reaction times during sleep deprivation. Behav Res Methods. 2017 Aug;49(4):1460-1469. doi: 10.3758/s13428-016-0802-5. PMID: 27631990.
\bibitem{sleep2peakapp}https://sleep-2-peak.com
\bibitem{sss1}"Management of Excessive Daytime Sleepiness Reviewed". Medscape. Retrieved 2021-04-07
\bibitem{s2papp}Brunet, JF., Dagenais, D., Therrien, M. et al. Validation of sleep-2-Peak: A smartphone application that can detect fatigue-related changes in reaction times during sleep deprivation. Behav Res 49, 1460–1469 (2017). https://doi.org/10.3758/s13428-016-0802-5
\bibitem{luca20}De Luca, Valeria, Luo, Hongyu, \& Clay, Ieuan. (2020). Continuous multi-sensor wearable data and daily subject-reported fatigue of heathy adults [Data set]. Zenodo. http://doi.org/10.5281/zenodo.4266157
\bibitem{luo20}Luo H., et. al. (2020), Assessment of Fatigue Using Wearable Sensors: A Pilot Study. Digit Biomark.
\bibitem{biovotion}https://www.biovotion.com/everion/
\bibitem{mich03}Michielsen HJ, De Vries J, Van Heck GL. Psychometric qualities of a brief self-rated fatigue measure: The Fatigue Assessment Scale. J Psychosom Res. 2003 Apr;54(4):345–52.
\bibitem{lee91}Lee KA, Hicks G, Nino-Murcia G. Validity and reliability of a scale to assess fatigue. Psychiatry Res. 1991 Mar;36(3):291–8.
\bibitem{cao18}Cao W, Wang D, Li J, Zhou H, Li L, Li Y. Brits: Bidirectional recurrent imputation for time series. In Advances in Neural Information Processing Systems. 2018. pp. 6775–85.
\bibitem{facenet}Florian Schroff, Dmitry Kalenichenko, James Philbin: FaceNet: A unified embedding for face recognition and clustering. CVPR 2015: 815-823
\bibitem{inception}C. Szegedy, W. Liu, Y. Jia, P. Sermanet, S. Reed, D. Anguelov, D. Erhan, V. Vanhoucke, and A. Rabinovich. Going deeper with convolutions. CoRR, abs/1409.4842, 2014. 
\bibitem{wtalc}Sujoy Paul, Sourya Roy, Amit K. Roy-Chowdhury: W-TALC: Weakly-Supervised Temporal Activity Localization and Classification. ECCV (4) 2018: 588-607
\bibitem{breath1}Haythem Rehouma, Rita Noumeir, Sandrine Essouri, Philippe Jouvet:
Quantitative Assessment of Spontaneous Breathing in Children: Evaluation of a Depth Camera System. IEEE Trans. Instrum. Meas. 69(7): 4955-4967 (2020)
\bibitem{breath2}Nir Regev, Dov Wulich: A simple, remote, video based breathing monitor. EMBC 2017: 1788-1791
\bibitem{act1}Vincent Jacquot, Zhuofan Ying, Gabriel Kreiman: Can Deep Learning Recognize Subtle Human Activities? CVPR 2020: 14232-14241
\bibitem{act2}Junwei Liang, Lu Jiang, Juan Carlos Niebles, Alexander G. Hauptmann, Li Fei-Fei: Peeking Into the Future: Predicting Future Person Activities and Locations in Videos. CVPR 2019: 5725-5734
\bibitem{rppg1}Yiming Liu, Binjie Qin, Rong Li, Xintong Li, Anqi Huang, Haifeng Liu, Yisong Lv, Min Liu: Motion-Robust Multimodal Heart Rate Estimation Using BCG Fused Remote-PPG With Deep Facial ROI Tracker and Pose Constrained Kalman Filter. IEEE Trans. Instrum. Meas. 70: 1-15 (2021)
\bibitem{rppg2}Giuseppe Boccignone, Donatello Conte, Vittorio Cuculo, Alessandro D'Amelio, Giuliano Grossi, Raffaella Lanzarotti:
An Open Framework for Remote-PPG Methods and Their Assessment. IEEE Access 8: 216083-216103 (2020)
\bibitem{rppg3}Wenjin Wang, Albertus C. den Brinker: Modified RGB Cameras for Infrared Remote-PPG. IEEE Trans. Biomed. Eng. 67(10): 2893-2904 (2020)
\bibitem{auth_ppg1}Turky N. Alotaiby, Fatima Aljabarti, Gaseb Alotibi, Saleh A. Alshebeili, Davide Palumbo: A Nonfiducial PPG-Based Subject Authentication Approach Using the Statistical Features of DWT-Based Filtered Signals. J. Sensors 2020: 8849845:1-8849845:14 (2020)
\bibitem{auth_ppg2}Yetong Cao, Qian Zhang, Fan Li, Song Yang, Yu Wang: PPGPass: Nonintrusive and Secure Mobile Two-Factor Authentication via Wearables. INFOCOM 2020: 1917-1926
\bibitem{auth_ppg3}Tianming Zhao, Yan Wang, Jian Liu, Yingying Chen: Your Heart Won't Lie: PPG-based Continuous Authentication on Wrist-worn Wearable Devices. MobiCom 2018: 783-785
\bibitem{acc_ppg1}Levi Benjamin Wood, H. Harry Asada: Noise Cancellation Model Validation for Reduced Motion Artifact Wearable PPG Sensors Using MEMS ACCs. EMBC 2006: 3525-3528
\bibitem{acc_ppg2}Alexander J. Casson, Arturo Vazquez Galvez, Delaram Jarchi: Gyroscope vs. ACC measurements of motion from wrist PPG during physical exercise. ICT Express 2(4): 175-179 (2016)
\bibitem{acc_ppg3}Delaram Jarchi, Alexander J. Casson: Description of a Database Containing Wrist PPG Signals Recorded during Physical Exercise with Both ACC and Gyroscope Measures of Motion. Data 2(1): 1 (2017)
\bibitem{acc_ppg4}	Giorgio Biagetti, Paolo Crippa, Laura Falaschetti, Simone Orcioni, Claudio Turchetti: Reduced complexity algorithm for heart rate monitoring from PPG signals using automatic activity intensity classifier. Biomed. Signal Process. Control. 52: 293-301 (2019)
\bibitem{acc_ppg5}	Anastasios Panagiotis Psathas, Antonios Papaleonidas, Lazaros Iliadis: Machine Learning Modeling of Human Activity Using PPG Signals. ICCCI 2020: 543-557
\bibitem{auth_eeg1}D. Farias-Castro, R. Salazar-Varas:
Person Authentication Based on Standard Deviation of EEG Signals and Bayesian Classifier. MICAI (2) 2020: 390-400
\bibitem{auth_eeg2}Eeva-Sofia Haukipuro, Ville Kolehmainen, Janne M. J. Huttunen, Sebastian Remander, Janne Salo, Tuomas Takko, Le Ngu Nguyen, Stephan Sigg, Rainhard Dieter Findling: Mobile Brainwaves: On the Interchangeability of Simple Authentication Tasks with Low-Cost, Single-Electrode EEG Devices. IEICE Trans. Commun. 102-B(4): 760-767 (2019)
\bibitem{hong20}Hongzhou Zhang, Peiyue Li, Zhiguo Du, Wei Dou: Risk Entropy Modeling of Surveillance Camera for Public Security Application. IEEE Access 8: 45343-45355 (2020)
\bibitem{abd20}Abdellah Chehri: Energy-efficient modified DCC-MAC protocol for IoT in e-health applications. Internet Things 14: 100119 (2021)
\bibitem{iot1} Udit Satija, Barathram Ramkumar, M. Sabarimalai Manikandan: Real-Time Signal Quality-Aware ECG Telemetry System for IoT-Based Health Care Monitoring. IEEE Internet Things J. 4(3): 815-823 (2017)
\bibitem{iot2}Prabal Verma, Sandeep K. Sood: Fog Assisted-IoT Enabled Patient Health Monitoring in Smart Homes. IEEE Internet Things J. 5(3): 1789-1796 (2018)
\bibitem{iot3}Mostafa Haghi, Sebastian Neubert, Andre Geissler, Heidi Fleischer, Norbert Stoll, Regina Stoll, Kerstin Thurow: A Flexible and Pervasive IoT-Based Healthcare Platform for Physiological and Environmental Parameters Monitoring. IEEE Internet Things J. 7(6): 5628-5647 (2020)
\bibitem{secure1}Ata Ullah, Muhammad Azeem, Humaira Ashraf, Abdulellah A. Alaboudi, Mamoona Humayun, N. Z. Jhanjhi: Secure Healthcare Data Aggregation and Transmission in IoT - A Survey. IEEE Access 9: 16849-16865 (2021)
\bibitem{secure2}Reyhane Attarian, Sattar Hashemi: An anonymity communication protocol for security and privacy of clients in IoT-based mobile health transactions. Comput. Networks 190: 107976 (2021)
\bibitem{secure3}Deepti Singh, Bijendra Kumar, Samayveer Singh, Satish Chand: A Secure IoT-Based Mutual Authentication for Healthcare Applications in Wireless Sensor Networks Using ECC. Int. J. Heal. Inf. Syst. Informatics 16(2): 21-48 (2021)
\bibitem{wtalc_code}https://github.com/sujoyp/wtalc-pytorch
\bibitem{activitynet}Fabian Caba Heilbron, Victor Escorcia, Bernard Ghanem, Juan Carlos Niebles: ActivityNet: A large-scale video benchmark for human activity understanding. CVPR 2015: 961-970
\bibitem{remote_ppg}	Pilz, C. S., Zaunseder, S., Krajewski, J., \& Blazek, V. (2018). Local group invariance for heart rate estimation from face videos in the wild. In Proceedings of the IEEE Conference on Computer Vision and Pattern Recognition Workshops (pp. 1254-1262).
\bibitem{remote_ppg_code} https://github.com/phuselab/pyVHR
\bibitem{wang17}Wang, L., Xiong, Y., Lin, D., Van Gool, L.: Untrimmednets for weakly supervised action recognition and detection. In: CVPR (2017)
\bibitem{imagenet}Deng, J., Dong, W., Socher, R., Li, L.J., Li, K., Fei-Fei, L.: Imagenet: A large-scale hierarchical image database. In: CVPR. pp. 248–255. IEEE (2009)
\bibitem{siamese}G. Koch, R. Zemel, R. Salakhutdinov, Siamese neural networks for one-shot image recognition, in: ICML Deep Learning Workshop, vol. 2, Lille, 2015.
\bibitem{gesture_keeper}Vasileios Sideridis, Andrew Zacharakis, George Tzagkarakis, Maria Papadopouli: GestureKeeper: Gesture Recognition for Controlling Devices in IoT Environments. CoRR abs/1903.06643 (2019)
\bibitem{gesture_keeper_code}https://github.com/guptajay/NUS-Hack-Roll-2020
\bibitem{encode_acc_code}https://github.com/lulujianjie/ACC-signal-recognition-using-ResNet-18-and-modified-Recurrence-Plots
\bibitem{encode_acc} Jianjie Lu and Raymond K. Y. Tong, Encoding ACC Signals as Images for Activity Recognition Using Residual Neural Network, https://arxiv.org/vc/arxiv/papers/1803/1803.09052v1.pdf
\bibitem{rp1}Marwan, N., Romano, M. C., Thiel, M., \& Kurths, J. Recurrence plots for the analysis of complex systems. Physics reports, 438(5-6), 237-329
\bibitem{rp2}Marwan, N. A historical review of recurrence plots. The European Physical Journal Special Topics, 164(1), 3-12.
\bibitem{41}Y. Sun, Y. Chen, X. Wang, and X. Tang, “Deep learning face representation by joint identification-verification,” Proceedings of the Advances in Neural Information Processing Systems, pp. 1988–1996, December 2014, Montreal, Canada
\bibitem{hr_ppg}Tang Lei, Zhu Cai, Luo Hua: Training prediction and athlete heart rate measurement based on multi-channel PPG signal and SVM algorithm. J. Intell. Fuzzy Syst. 40(4): 7497-7508 (2021)
\bibitem{br_ppg}Vivek Chandel, Jayeeta Saha, Chirayata Bhattacharyya, Avik Ghose: Real-time robust estimation of breathing rate from PPG using commercial-grade smart devices: demo abstract. SenSys 2020: 633-634
\bibitem{eda_feat}Md. Rafiul Amin, Rose T. Faghih: Tonic and Phasic Decomposition of Skin Conductance Data: A Generalized-Cross-Validation-Based Block Coordinate Descent Approach. EMBC 2019: 745-749
\bibitem{mSNN}Debashis Das Chakladar, Pradeep Kumar, Partha Pratim Roy, Debi Prosad Dogra, Erik J. Scheme, Victor Chang: A multimodal-Siamese Neural Network (mSNN) for person verification using signatures and EEG. Inf. Fusion 71: 17-27 (2021)

\end{thebibliography}
\end{document}